\documentclass{article-hermes}
\usepackage{graphicx}

\begin{document}
\proceedings{GEOX2010}{1}
\title[Numerical Modeling of Porous Media]%
      {Numerical Modeling of Complex Porous Media For Borehole Applications}
\subtitle{NMR-response and Transport in Carbonate and Sandstone Rocks}
\author{%
   Seungoh Ryu\fup{*} \andauthor Weishu Zhao\fup{*} 
   \andauthor Gabriela Leu\fup{*} \andauthor Philip M. Singer\fup{**} \andauthor Hyung Joon Cho\fup{***}
   \andauthor Youngseuk Keehm\fup{****}
}
\address{
   \fup{*}Schlumberger Doll Research\\
Cambridge, MA,  02139, USA\\[3pt]
   \fup{**}Schlumberger Dhahran Carbonate Research\\
  Al-Khobar 31952, Kingdom of Saudi Arabia\\[6pt]
  \fup{***} Schlumberger Product Center\\
  Sugarland, TX, 77478, USA \\[3pt]
   \fup{****}Dept. of Geoenvironmental Sciences\\
   Kongju National University,  Kongju, South Korea\\[6pt]
}
\abstract{The diffusion/relaxation behavior of polarized spins of pore filling fluid, as often probed by NMR relaxometry, is widely used to extract information on the pore-geometry. Such information is further interpreted as an indicator of the key transport property of the formation in the oil industry. As the importance of reservoirs with complex pore geometry grows, so does the need for deeper understanding of how these properties are inter-related. Numerical modeling of relevant physical processes using a known pore geometry promises to be an effective tool in such endeavor.
Using a suite of numerical techniques based on random-walk (RW) and Lattice-Boltzmann (LB) algorithms, we compare sandstone  and carbonate pore geometries in their impact on NMR and flow properties. For NMR relaxometry, both laboratory measurement and simulation were done on the same source to address some of the long-standing issues in its borehole applications.
Through a series of "numerical experiments" in which the interfacial relaxation properties of the pore matrix is varied systematically, we study the effect of a variable surface relaxivity while fully incorporating the complexity of the pore geometry. From combined RW and LB simulations, we also obtain diffusion-convection propagator and compare the result with experimental and network-simulation counterparts.
 }
\keywords{NMR, Lattice-Boltzmann, Randomwalk, Tomogram, Carbonate, Sandstone Rocks}

\maketitlepage

\section{Introduction}
Despite the long history of research on porous media, there remain open issues which critically affect various industrial endeavor such as oil/gas exploration, ${\rm CO_2}$ sequestration, water management, storage and migration of toxic waste. A wide range of scales permeates through these disciplines, but pore-scale physical processes remain their common denominator. In this work, we report our recent effort on the pore-level modeling based on micro-tomograms in the context of a borehole application. 

\begin{figure}[htbp]
   \centering
   \includegraphics[width=3.8in]{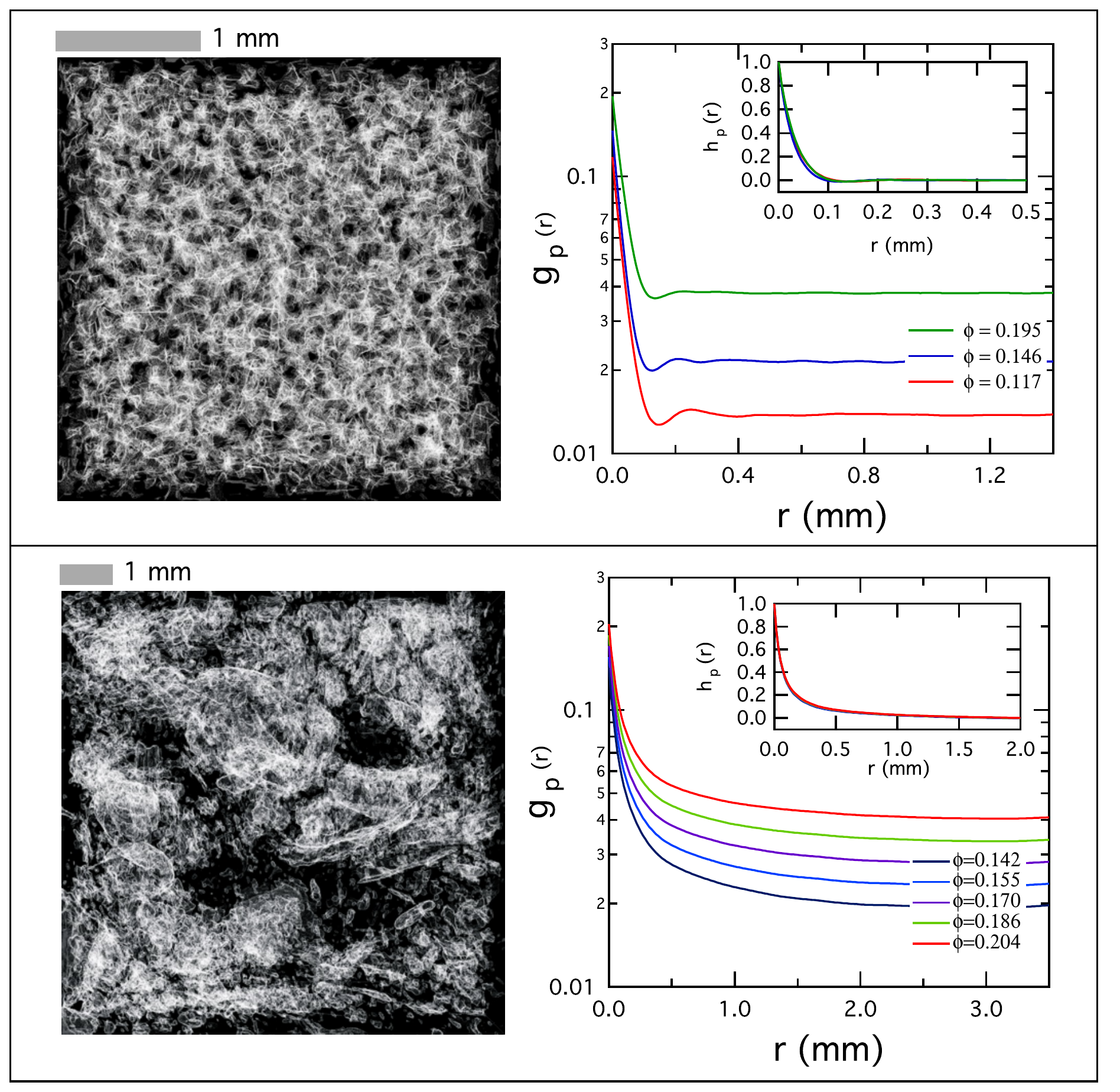}
    \caption{Pore-grain interface morphology for a typical sandstone (top) and a carbonate rock(bottom panel). Also shown are the radial pore-to-pore auto-correlation function $g_p(r)$ at various porosity values (top:three distinct sandstones; bottom:different thresholding of the original carbonate tomogram). The insets show scaled correlation functions $h_p(r)$ which largely collapse into a {\it universal} form.}
   \label{fig:comparepore}
\end{figure}

Several techniques (e.g.  \cite{Auzerais:1996p660}) utilizing detailed 3D pore geometry have reached maturity in recent decade thanks to the affordable computing resource, imaging techniques,  and parallelized simulations. These numerical results are in good standing for a class of porous media such as bead packs and sandstones. Here, we focus on aspects of more challenging situations involving carbonates  in the oil field.

\begin{figure}[htbp] 
   \centering
   \includegraphics[width=4.in]{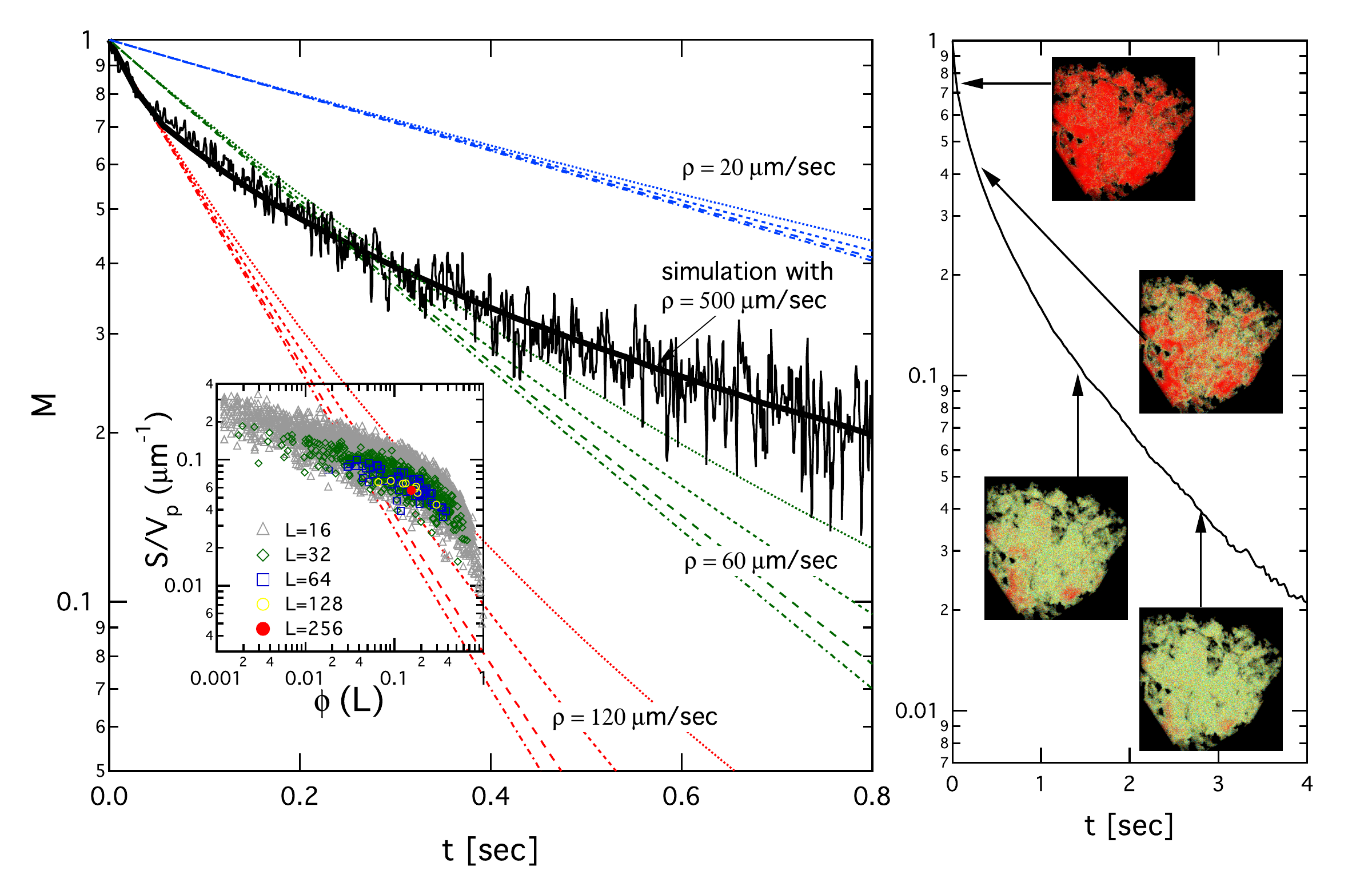}
   \caption{Low-field experimental data (solid zagged curve) and the best simulation match (solid curve) with $\rho = 500 {\rm \mu m/s}$. Simplistic model results for NMR response at various strengths of surface relaxivity $\rho$ and pore coarsening (colored broken lines, four curves for each value of $\rho$) are also shown for comparison. The inset shows the porosity and $S/V_p$ distribution at the five stages of coarse graining. 
Shown on the right panel are the local magnetization density evolution at various stages of the simulation for $\rho = 500 \mu {\rm m/s}$ which yields results matching the experiment.}
   \label{fig:nmrall}
\end{figure}

\section{Pore geometry and open issues for the carbonate}
Figure \ref{fig:comparepore} shows two contrasting images of the pore-grain interface for a Fontainebleu sandstone and a carbonate rock. While quasi-periodicity is clearly visible in the former, the latter displays pronounced heterogeneity. This is quantified in the right column where the radial pore-to-pore autocorrelation function $g(|{\bf r}_2 - {\bf r}_1|) \equiv < \phi({\bf r}_1) \phi({\bf r}_2) >$ ($\phi ({\bf r}) = 1$ in pore, $0$ in grains) is plotted for different values of porosity. Note that in both cases, the curves collapses to a generic form (insets) $h_p(r) \equiv (g(r) - \phi^2)/ ( \phi ( 1 - \phi))$ with $\phi = <\phi ({\bf r})>$, average porosity. Quasi-periodicity of the sandstone is manifest in the form of mild bumps in $g(r)$, reminiscent of the classic density correlation in a simple liquid, while the carbonate sample displays a monotonic, featureless profile, which actually arises from preponderance of many competing length scales. In many carbonate rocks, the complexity extends even further than suggested in the figure, as there exist structures on finer scales beyond the tomographic resolution, as well as on scales much beyond the typical sample sizes. Quantitative elucidation of both these aspects remain an open challenge which we aim to address via further extending the steps described in the following.

\begin{figure}[htbp] 
   \centering
   \includegraphics[width=3.6in]{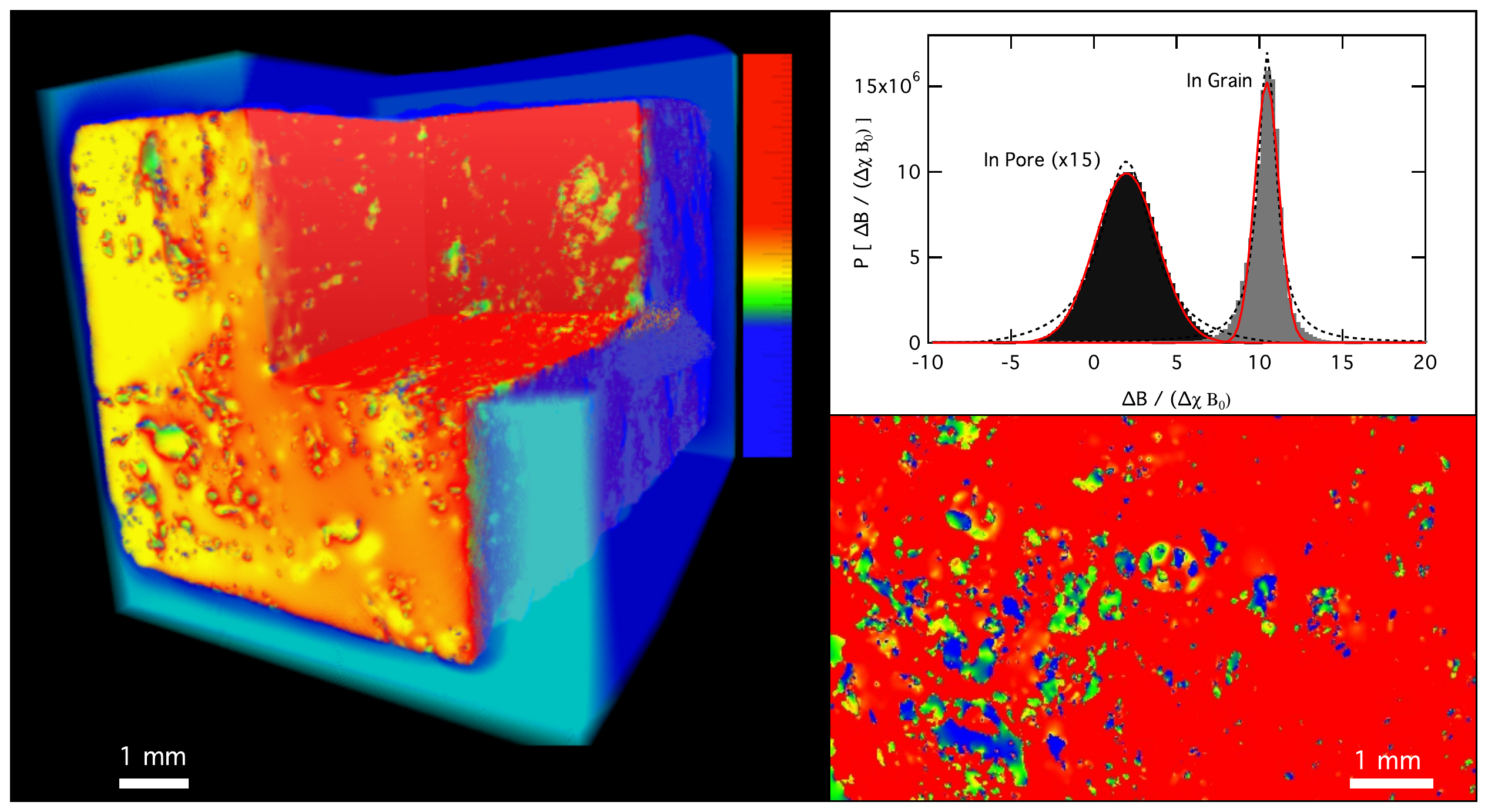} 
   \caption{The internal magnetic field of a carbonate packstone numerically obtained for the entire tomogram ($1.5^3 {\rm cm}^3$) of the carbonate rock. Shown on the right are the slice-cut views of the field and the probability distribution of its strength for a sub-block of $512^3$ voxels at the center rendered separately for the pore (filled black) and grain (gray) space. The curves represent the best Lorentzian (broken lines)\cite{Chen:2005p598} and Gaussian (solid lines) fits. The latter works better for the pore portion, while neither of the methods successfully fits the grain portion.}
   \label{fig:internalfield}
\end{figure}

\section{Numerical NMR relaxometry}
Simulations based on realistic 3D pore allows us to address some of the long-standing issues in probing its geometry. One surrounds the validity of the conventional mapping between the NMR relaxation spectrum and the pore-size distribution\cite{Kleinberg:1999p868,Grebenkov:2007p774} of carbonate rocks in the possible presence of haphazard heterogeneity in their interfacial properties.\cite{Ryu:2009p753}  The {\em low field} NMR data of the carbonate sample of Fig.\ref{fig:comparepore} (black zagged line in Fig.\ref{fig:nmrall} for which bulk relaxation rate of $1/T_b = 0.67 {\rm s}^{-1}$ was used and 5-points moving average was taken), shows the typical multi-exponential characteristics. This often invites an interpretation in terms of a broad pore size (more precisely, the surface-to-volume (S-V) ratios) distribution. In this scenario, the pore space is approximated as a collection of {\em isolated} pores of varying sizes, all in the so-called {\em fast diffusion} limit\cite{Brownstein:1979p779}. From the tomogram of the same piece, we obtain the distribution ${\cal P}$ of local $\{ S_i/V_i \}$ and porosity  $\{V_i \}$ (shown in the inset of Figure \ref{fig:nmrall}) at various stages of coarse-graining ($i$ labeling sub-blocks of linear dimension $L$). Attempts to fit experimental data using such recipe neglect the diffusive coupling between pores as shown by series of curves in the figure with a range of controlling surface relaxivity parameter, $\rho$, values \cite{Kleinberg:1999p868} (broken lines with $\rho = 20, 60, 120 \mu m/s$) and they fail to work: if $\rho$ is chosen to match the initial slope (red curves), it fails to match the experimental data at long times, and vice versa (blue lines). This suggests that the model neglecting the extended nature of the pore and the heterogeneous diffusive-coupling among its constituents has limited validity in these types of rocks. 
The role of the latter had been previously considered in simple 1D models.\cite{Ramakrishnan:1999p995,Zielinski:2002p769}
Random-walk based simulation\cite{Ryu:2008p531} on the 3D tomographic pore yields an excellent overall agreement (solid black curve) with the choice of $500 \mu m/s$ for the single parameter. The large $\rho$ value thus inferred partly accounts for the fact that at the resolution of $17 \mu m$ per voxel, the digital representation of the interface significantly under-estimates the surface area. 
The effective $\tilde\rho$, which represents the combination of raw $\rho$ value and the actual surface-area, dictates the long time scale dynamics responsible for the good agreement with the entirety of data. 
 
These numerics can be extended for sophisticated NMR probes\cite{Song:2000p543, Song:2008p724}. The internal field arising from the weak susceptibility contrast between the rock matrix and fluid, often a nuisance for NMR probes, offers a way to probe length scales thanks to the close geometrical correlation between its spatial profile and the geometry of the matrix.\cite{Song:2000p543} Figure \ref{fig:internalfield} shows an example of the internal field profile in the carbonate rock calculated using a weak dipole field {\em ansatz}, a method verified via direct imaging on a pack of cylindrical tubes.\cite{Cho:2009p776} We further derive the local field gradients that play a critical role in stimulated-echo probes\cite{Song:2000p543} as well as NMR at high fields.\cite{Anand:2007p741,Ryu:2009p986} We find that the field inside the 3D pore space can be better  approximated by a Gaussian distribution rather than a Lorenzian as reported by Chen {\em et al}.\cite{Chen:2005p598} A detailed study on the statistics of internal field and their gradients inside various types of rocks is under way, as well as their effect on the NMR response. 

\begin{figure}[htbp] 
   \centering
   \includegraphics[width=3.6in]{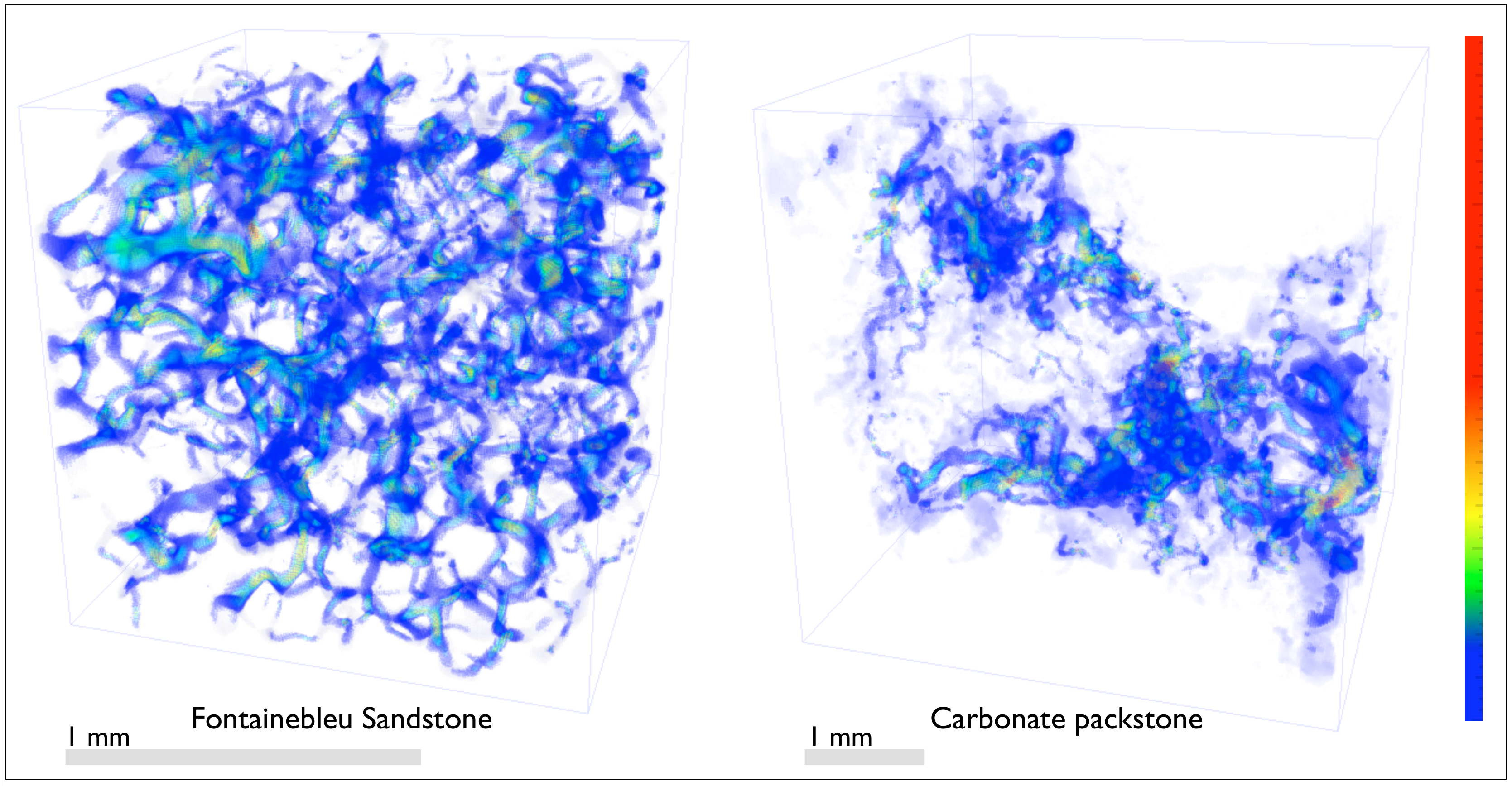}
      \caption{Local flow speed through the pore matrix (Fontainebleu sandstone (left) and a carbonate packstone (right)), driven from left to right. The color represents the local speed and its scheme was chosen to enhance its features. In both cases, the voxels with less than 2 \% of the maximum speed are omitted from view for enhanced views. }
   \label{fig:flowspeeds}
\end{figure}

\section{Diffusion-Flow propagators}
One of the objectives of our numerical modeling is to improve the link between the static pore geometry (as inferred {\em in situ} from borehole measurements) and its transport, since the latter ultimately controls viability of a hydrocarbon reservoir. 
The stark contrast between the sandstone and the carbonate pore geometry (Fig.\ref{fig:comparepore}) underlies the distinct way how the flow is distributed. Figure \ref{fig:flowspeeds} shows the local flow speed distribution for the sub-volume of the rocks obtained from the Lattice Boltzmann simulations.\cite{Succi:2001p780} Clearly, the inhomogeneous flux distribution and extreme tortuosity as apparent in the carbonate sample hints at why any framework based soley on parallel channels of effective hydraulic radii should fail. 

\begin{figure}[htbp] 
   \centering
   \includegraphics[width=4in]{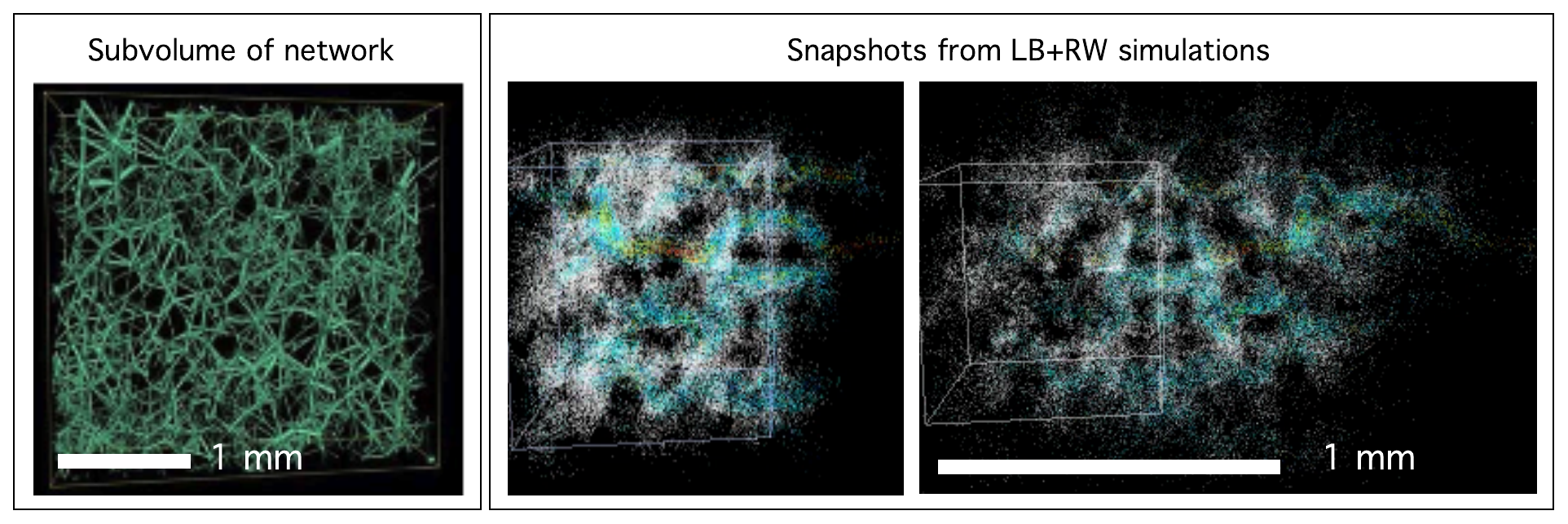}
      \caption{The first panel shows the carbonate rock as used in the network simulation. The second panel shows the random-walk-LB based propagator simulation on the sub volume of the same rock at two different stages of mean displacement. }
   \label{fig:poreviews}
\end{figure}
\begin{figure}[htbp] 
   \centering
   \includegraphics[width=4in]{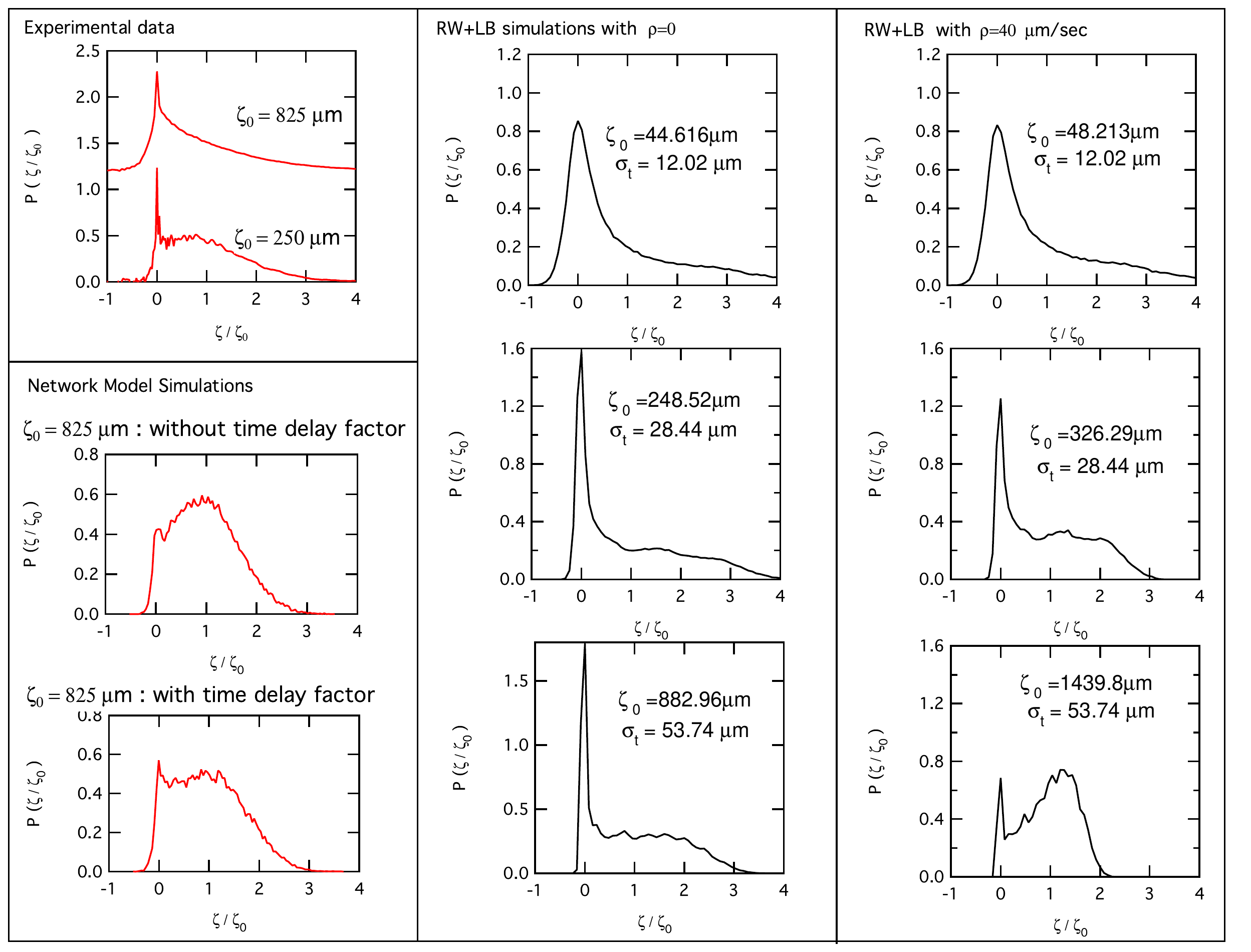}
      \caption{Top panel on the first column shows experimental $P(\zeta,\zeta_0)$ at two different stages. The lower panels show the result of a network model simulation {\em with} and {\em without} the {\em ad hoc} time dealy. The middle column shows the combined RW-LB simulated propagator with $\rho=0$ at three different stages (parametrized by the average displacement $\zeta_0$). The third column shows the same but with $\rho = 40 {\rm \mu m/s}$. The actual $\rho$, from more recent NMR relaxometry, is estimated to be about $20{\rm \mu m/s}$. Also indicated for reference is the diffusion length $\sigma_t \equiv \sqrt{2 D t}$ for each stage.}
   \label{fig:props}
\end{figure}

The flow propagator\cite{deGennes:1983p727,Stanley:1984p987,Aronovitz:1984p548} which stands for the probability distribution of tagged particle displacements at two different wait-times (as indicated by the average displacement $\zeta_0$), $P(\zeta, \zeta_0)$ provides  further details on the interplay between the geometrical restriction and the flow, and can be measured by NMR\cite{Hulin:1991p557, Lebon:1996p620,Scheven:2004p474}. We simulate its process using two complementary techniques: one based on network reduction, the other through a combination of random walk and lattice-Boltzmann. (See Figure \ref{fig:poreviews}) The former addresses much bigger volumes, while the latter incorporates detailed diffusion/fluid dynamics at finer length scales. The top-left panel of Figure \ref{fig:props} shows the experimental data on a {\em clean} dolomite rock. (The rock has a much simpler pore geometry than the carbonate rock used in Figs.1-4.) From the tomogram of the same source rock, we first ran a network-model simulation.\cite{Zhao:2009p807}  The result (lower-left panels) under-represents the sharp peak near the origin present in the experimental data. An {\em ad-hoc} time delay factor was introduced, which to a certain degree enhances the weight near zero displacement\cite{Zhao:2009p807}. Questions arise as to what degree one requires the presence of sub-micron-pores which lie beyond the resolution of the tomogram, and how large-scale heterogeneity and a finite $\rho$ affect $P(\zeta, \zeta_0)$.
To clarify these, we developed an alternative method by allowing the random walkers to ride on the background flow from an LB run. (the second panel of Fig.\ref{fig:poreviews}) Preliminary results (the 2nd and 3rd columns for $\rho=0$ and $\rho=40\mu m/s$) based on the resolution $\sim 3 \mu m/vox$ seem to capture the most salient features of the data. 
This improved agreement provides a valuable insight on how one should attribute such experimental features to the heterogeneity at extreme length scales. More quantitative inquiry into this issue is in progress.
 
The authors would like to thank Drs. J. Dunsmuir (Brookhaven NL), J. Goebbels (BAM) and M. Knackstedt (ANU) for microtomograms of various rocks.
\bibliography{sryu_geox10}

\begin{thebibliography}{}

\bibitem[ANA~07]{Anand:2007p741}
\textsc{Anand V.}\andname{}\textsc{Hirasaki G.}, \guilo{}Paramagnetic
  relaxation in sandstones: distringuishing T1 and T2 dependence on surface
  relaxation, internal gradients and dependence on echo spacing\guilf{},
\newblock \textit{SPWLA 48th Annual Logging Symposium}, \volumename\ 446359V,
  2007.

\bibitem[ARO~84]{Aronovitz:1984p548}
\textsc{Aronovitz J.}\andname{}\textsc{Nelson D.}, \guilo{}Anomalous diffusion
  in steady fluid flow through a porous medium\guilf{},
\newblock \textit{Phys. Rev. A}, \volumename\ 30, \numbername\ 4, 1984,
  \pagename{} 1948.

\bibitem[AUZ~96]{Auzerais:1996p660}
\textsc{Auzerais F.}, \textsc{Dunsmuir J.}, \textsc{Ferreol B.}, \textsc{Martys
  N.}, \textsc{Olson J.}, \textsc{Ramakrishnan T.}, \textsc{Rothman
  D.}\andname{}\textsc{Schwarts L.}, \guilo{}Transport in sandstone: A study
  based on three dimensional microtomography\guilf{},
\newblock \textit{Geo. Phys. Lett.}, \volumename\ 23, 1996,  \pagename{} 705.

\bibitem[BRO~79]{Brownstein:1979p779}
\textsc{Brownstein K.}\andname{}\textsc{Tarr C.}, \guilo{}Importance of
  classical diffusion in NMR studies of water in biological cells\guilf{},
\newblock \textit{Phys. Rev. A}, \volumename\ 19, 1979,  \pagename{} 2446.

\bibitem[CHE~05]{Chen:2005p598}
\textsc{Chen Q.}, \textsc{Marble A.}, \textsc{Colpitts
  B.}\andname{}\textsc{Balcom B.}, \guilo{}The internal magnetic field
  distribution, and single exponential magnetic resonance free induction decay,
  in rocks\guilf{},
\newblock \textit{Journal of Magnetic Resonance}, \volumename\ 175,
  \numbername\ 2, 2005,  \pagesname{} 300--308.

\bibitem[CHO~09]{Cho:2009p776}
\textsc{Cho H.}, \textsc{Ryu S.}, \textsc{Ackerman J.}\andname{}\textsc{Song
  Y.-Q.}, \guilo{}Visualization of inhomogeneous local magnetic field gradient
  due to susceptibility contrast\guilf{},
\newblock \textit{J. Mag. Res.}, \volumename\ 198, 2009,  \pagename{}~88.

\bibitem[GEN~83]{deGennes:1983p727}
\textsc{de~Gennes P.}, \guilo{}Hydrodynamic dispersion in unsaturated porous
  media\guilf{},
\newblock \textit{Journal of Fluid Mechanics Digital Archive}, \volumename\
  136, \numbername\ 1, 1983,  \pagesname{} 189--200,
\newblock 10.1017/S0022112083002116.

\bibitem[GRE~07]{Grebenkov:2007p774}
\textsc{Grebenkov D.}, \guilo{}NMR survey of reflected Brownian motion\guilf{},
\newblock \textit{Rev. Mod. Phys.}, \volumename\ 79, 2007,  \pagesname{}
  1077--61.

\bibitem[HUL~91]{Hulin:1991p557}
\textsc{Hulin J.}, \textsc{Guyon E.}, \textsc{Charlaix E.}, \textsc{Leroy
  C.}\andname{}\textsc{Magnico P.}, \guilo{}Abnormal diffusion and dispersion
  in porous media\guilf{},
\newblock \textit{Physica Scripta}, \volumename\ 1991, \numbername\ T35, 1991,
  \pagename{}~26.

\bibitem[KLE~99]{Kleinberg:1999p868}
\textsc{Kleinberg R.}, \guilo{}Methods in the physics of porous media\guilf{},
\newblock \Inname{} \textsc{Wong P.}, \editorname{}, \textit{Nuclear Magnetic
  Resonance}, \volumename{}~35, Academic Press, 1999.

\bibitem[LEB~96]{Lebon:1996p620}
\textsc{Lebon L.}, \textsc{Oger L.}, \textsc{Leblond J.}, \textsc{Hulin J.},
  \textsc{Martys N.}\andname{}\textsc{Schwartz L.}, \guilo{}Pulsed gradient NMR
  measurements and numerical simulation of flow velocity distribution in sphere
  packings\guilf{},
\newblock \textit{Physics of Fluids Physics of Fluids Phys. Fluids},
  \volumename\ 8, \numbername\ 2, 1996,  \pagesname{} 293--301.

\bibitem[RAM~99]{Ramakrishnan:1999p995}
\textsc{Ramakrishnan T.~S.}, \textsc{Schwartz L.~M.}, \textsc{Fordham E.~J.},
  \textsc{Kenyon W.~E.}\andname{}\textsc{Wilkinson D.~J.}, \guilo{}Forward
  Models for Nuclear Magnetic Resonance in Carbonate Rocks\guilf{},
\newblock \textit{The Log Analysist}, \volumename\ 40, 1999,  \pagename{} 260.

\bibitem[RYU~08]{Ryu:2008p531}
\textsc{Ryu S.}, \guilo{}Effects of Spatially Varying Surface Relaxivity and
  Pore Shape on NMR Logging\guilf{},
\newblock \textit{SPWLA Proceedings of the 49th Annual Logging Symposium,
  SPWLA}, , 2008,  \pagename{} 737008 BB.

\bibitem[RYU~09a]{Ryu:2009p986}
\textsc{Ryu S.}, \guilo{}Effect of inhomogeneous surface relaxivity, pore
  geometry and internal field gradient on NMR logging : exact and perturbative
  theories and numerical investigations\guilf{},
\newblock \textit{SPWLA 2009}, , 2009,  \pagename{} JJJJ.

\bibitem[RYU~09b]{Ryu:2009p753}
\textsc{Ryu S.}\andname{}\textsc{Johnson D.}, \guilo{}Aspects of
  diffusive-relaxation dynamics with a {\it non-uniform}, partially absorbing
  boundary in general porous media\guilf{},
\newblock \textit{Phys. Rev. Lett.}, \volumename\ in press, 2009.

\bibitem[SCH~04]{Scheven:2004p474}
\textsc{Scheven U.}, \textsc{Seland J.}\andname{}\textsc{Cory D.}, \guilo{}NMR
  propagator measurements on flow through a random pack of porous glass beads
  and how they are affected by dispersion, relaxation, and internal field
  inhomogeneities\guilf{},
\newblock \textit{Phys. Rev. E}, \volumename\ 69, 2004,  \pagename{} 021201.

\bibitem[SON~00]{Song:2000p543}
\textsc{Song Y.-Q.}, \textsc{Ryu S.}\andname{}\textsc{Sen P.},
  \guilo{}Determining multiple length scales in rocks\guilf{},
\newblock \textit{Nature}, \volumename\ 406, 2000,  \pagename{} 178.

\bibitem[SON~08]{Song:2008p724}
\textsc{Song Y.-Q.}, \textsc{Zielinski L.}\andname{}\textsc{Ryu S.},
  \guilo{}Two-Dimensional NMR of Diffusion Systems\guilf{},
\newblock \textit{Phys. Rev. Lett.}, \volumename\ 100, 2008,  \pagename{}
  248002.

\bibitem[STA~84]{Stanley:1984p987}
\textsc{Stanley H.~E.}\andname{}\textsc{Coniglio A.}, \guilo{}Flow in Porous
  Media: The `Backbone' Fractal at the Percolation Threshold\guilf{},
\newblock \textit{Physical Review B}, \volumename\ 29, 1984,  \pagename{} 522.

\bibitem[SUC~01]{Succi:2001p780}
\textsc{Succi S.}, \textit{The Lattice Boltzmann Equation for Fluid Dynamics
  and Beyond},
\newblock Oxford University Press, 2001.

\bibitem[ZHA~09]{Zhao:2009p807}
\textsc{Zhao W.}, \textsc{Picard G.}, \textsc{Leu G.}\andname{}\textsc{Singer
  P.}, \guilo{}Characterization of single phase flow through carbonate rocks:
  Quantitative comparison of NMR flow propagator measurements with a realistic
  pore network model\guilf{},
\newblock \textit{Trans. Porous. Media}, , 2009.

\bibitem[ZIE~02]{Zielinski:2002p769}
\textsc{Zielinski L.}, \textsc{Song Y.-Q.}, \textsc{Ryu
  S.}\andname{}\textsc{Sen P.}, \guilo{}Characterization of coupled pore
  systems from the diffusion eigenspectrum\guilf{},
\newblock \textit{J. of Chem. Phys.}, \volumename\ 117, \numbername\ 11, 2002,
  \pagesname{} 5361--5365.

\end{thebibliography}
\end{document}